# Defining the scope of AI regulations

*Jonas Schuett**

The paper argues that the material scope of AI regulations should not rely on the term 'artificial intelligence (AI)'. The argument is developed by proposing a number of requirements for legal definitions, surveying existing AI definitions, and then discussing the extent to which they meet the proposed requirements. It is shown that existing definitions of AI do not meet the most important requirements for legal definitions. Next, the paper argues that a risk-based approach would be preferable. Rather than using the term AI, policy makers should focus on the specific risks they want to reduce. It is shown that the requirements for legal definitions can be better met by defining the main sources of relevant risks: certain technical approaches (e.g. reinforcement learning), applications (e.g. facial recognition), and capabilities (e.g. the ability to physically interact with the environment). Finally, the paper discusses the extent to which this approach can also be applied to more advanced AI systems.

## 1. Introduction

Policy makers around the world are currently working on AI regulations.[1] In 2021, the European Commission published a proposal for an Artificial Intelligence Act (AI Act),[2] which is generally seen as the first comprehensive attempt to regulate AI in a major jurisdiction. The US has been more hesitant so far.

---

* Research Fellow, Centre for the Governance of AI, Oxford, UK; Research Affiliate, Legal Priorities Project, Cambridge, MA, USA; PhD Candidate, Faculty of Law, Goethe University Frankfurt, Germany; jonas.schuett@governance.ai.

[1] By 'regulation', I mean 'sustained and focused attempts to change the behaviour of others in order to address a collective problem or attain an identified end or ends, usually but not always through a combination of rules or norms and some means for their implementation and enforcement, which can be legal or non-legal', Julia Black and Andrew D Murray, 'Regulating AI and Machine Learning: Setting the Regulatory Agenda' (2019) 10 *European Journal of Law and Technology*, https://perma.cc/A456-QPHH; see also Christel Koop and Martin Lodge, 'What is Regulation? An Interdisciplinary Concept Analysis' (2017) 11 *Regulation and Governance* 95, https://doi.org/10.1111/rego.12094.

[2] European Commission, 'Proposal for a Regulation Laying Down Harmonised Rules on Artificial Intelligence (Artificial Intelligence Act)' COM (2021) 206 final, https://perma.cc/V2RH-6KGC.

Under the Trump administration, the focus was more on removing regulatory barriers,[3] but this focus has shifted under the Biden administration. Although there has been some work at the federal level,[4] most efforts to regulate AI seem to take place at the state level.[5] China's approach to AI regulation has also changed over the past two years. Initially, the focus was on voluntary AI ethics principles similar to those published by Western institutions.[6] But since 2020, more stringent regulations for AI companies were introduced, while state use of AI remains completely unrestricted.[7] This global dynamic has already been framed as a 'race to regulate AI'.[8]

One challenge faced by all policy makers who work on AI regulation is how to define the scope of application, which determines whether or not a regulation is applicable in a particular case. The scope of application defines *what* is regulated (material scope), *who* is regulated (personal scope), *where* the regulation applies (territorial scope), and *when* it applies (temporal scope). In this paper, I focus on the material scope. The territorial and temporal scope depend on jurisdiction-specific details, and defining the personal scope is a difficult question which deserves a paper on its own. The scope of application is described in the body of the regulation, using terms typically defined elsewhere in the regulation. These definitions are called legal definitions. The distinction between the terms that are used to define the scope of application ('this regulation applies to AI') and the definitions of these terms ('AI means…') will be important throughout this paper because the core argument is based on the conjunction between the two ('policy makers should only use the term AI for the scope definition if there is a good definition of AI').

Defining the scope of AI regulations is particularly challenging because the term AI is used for so many different systems—'it isn't any one thing'.[9] It can refer to systems that play games,[10] produce coherent text,[11] predict protein structures,[12] diagnose eye diseases,[13] or control nuclear fusion reactors.[14] From a regulatory perspective, these systems have very different risk profiles and therefore must be treated differently. To further complicate things, the term AI is highly ambiguous. There is a vast spectrum of definitions,[15] and its meaning changes over time. As famously put by John McCarthy: 'as soon as it works, no one calls it AI any more'.[16]

---

[3] The White House, 'Maintaining American Leadership in Artificial Intelligence' (2019) Executive Order 13859, https://perma.cc/MAN8-7TJJ.

[4] E.g. Eric Lander and Alondra Nelson, 'Americans Need a Bill of Rights for an AI-Powered World' (*The White House*, 22 October 2021) https://perma.cc/6ZRX-Q9ZB.

[5] See National Conference of State Legislatures, 'Legislation Related to Artificial Intelligence' (2022) https://perma.cc/49NS-WE9Y.

[6] E.g. Beijing Academy of Artificial Intelligence, 'Beijing AI Principles' (2019) https://perma.cc/PHA3-NUGY.

[7] Jennifer Conrad and Will Knight, 'China Is About to Regulate AI—and the World Is Watching' (2022) https://perma.cc/6ACT-WW4M.

[8] Nathalie A Smuha, 'From a "Race to AI" to a "Race to AI Regulation": Regulatory Competition for Artificial Intelligence' (2021) 13 *Law, Innovation and Technology* 57, https://doi.org/10.1080/17579961.2021.1898300.

[9] Peter Stone and others, 'Artificial Intelligence and Life in 2030' (*Stanford University*, 2016) https://perma.cc/36VX-Y6MM, 48.

The question of how to define AI in legal terms—especially in a regulatory context—has been raised by many legal scholars. While some have suggested the need for a single legal definition of AI,[17] others have argued that this is not feasible.[18] However, there are three notable gaps in the current literature. First, although most arguments rely on certain requirements for legal definitions (e.g. being future-proof), there seems to be no meta-discussion about these requirements. They tend to be treated as something given, without any justification of their legal origin or appropriateness. Second, there is no comprehensive discussion of all requirements; different scholars focus on different requirements. Third, there is only limited discussion of alternative approaches.

---

[10] E.g. Oriol Vinyals and others, 'Grandmaster Level in StarCraft II Using Multi-Agent Reinforcement Learning' (2019) 575 *Nature* 350, https://doi.org/10.1038/s41586-019-1724-z; Julian Schrittwieser and others, 'Mastering Atari, Go, Chess and Shogi by Planning with a Learned Model' (2020) 588 *Nature* 604, https://doi.org/10.1038/s41586-020-03051-4; OpenAI and others, 'Dota 2 with Large Scale Deep Reinforcement Learning' (2019) https://arxiv.org/abs/1912.06680.

[11] E.g. Jacob Devlin and others, 'BERT: Pre-training of Deep Bidirectional Transformers for Language Understanding' (2018) https://arxiv.org/abs/1810.04805; Tom B Brown and others, 'Language Models are Few-Shot Learners' (2020) https://arxiv.org/abs/2005.14165; Jack W Rae and others, 'Scaling Language Models: Methods, Analysis & Insights from Training Gopher' (2021) https://arxiv.org/abs/2112.11446; Jordan Hoffmann and others, 'Training Compute-Optimal Large Language Models' (2022) https://arxiv.org/abs/2203.15556; Aakanksha Chowdhery and others, 'PaLM: Scaling Language Modeling with Pathways' (2022) https://arxiv.org/abs/2204.02311.

[12] E.g. Andrew W Senior and others, 'Improved Protein Structure Prediction Using Potentials from Deep Learning' (2020) 577 *Nature* 706, https://doi.org/10.1038/s41586-019-1923-7; Kathryn Tunyasuvunakool and others, 'Highly Accurate Protein Structure Prediction for the Human Proteome' (2021) 596 *Nature* 590, https://doi.org/10.1038/s41586-021-03828-1.

[13] E.g. Jason Yim and others, 'Predicting Conversion to Wet Age-Related Macular Degeneration Using Deep Learning' (2020) 26 *Nature Medicine* 892, https://doi.org/10.1038/s41591-020-0867-7.

[14] E.g. Jonas Degrave and others, 'Magnetic Control of Tokamak Plasmas through Deep Reinforcement Learning' (2022) 602 *Nature* 414, https://doi.org/10.1038/s41586-021-04301-9.

[15] Shane Legg and Marcus Hutter, 'A Collection of Definitions of Intelligence' (2007) https://arxiv.org/abs/0706.3639; Sofia Samoili and others, 'AI Watch: Defining Artificial Intelligence' (*European Commission*, 2020) https://doi.org/10.2760/382730.

[16] Bertrand Meyer, 'John McCarthy' (*Communications of the ACM*, 28 October 2011) https://perma.cc/49S8-3GM6.

[17] Gary Lea, 'Why We Need a Legal Definition of Artificial Intelligence' (*The Conversation*, 2 September 2015) https://perma.cc/6NZG-5KCS; Jacob Turner, *Robot Rules: Regulating Artificial Intelligence* (Palgrave Macmillan 2019) https://doi.org/10.1007/978-3-319-96235-1, 7-8; Rex Martinez, 'Artificial Intelligence: Distinguishing between Types & Definitions' (2019) 19 *Nevada Law Journal* 1015, https://perma.cc/F8YN-7RKZ, 1022.

The paper proceeds as follows. First, I argue that policy makers should not rely on the term AI to define the material scope of AI regulations. Next, I argue that policy makers should instead consider using certain technical approaches, applications, and capabilities, following a risk-based approach. Finally, I discuss the extent to which this approach can also be applied to more advanced AI systems.

## 2. Should policy makers use the term AI to define the material scope of AI regulations?

The most obvious way to define the material scope of AI regulations would be to use the term AI. For example, Article 2(1) of the AI Act uses the following formulation:

> This Regulation applies to (a) providers placing on the market or putting into service AI systems in the Union, irrespective of whether those providers are established within the Union or in a third country; (b) users of AI systems located within the Union; (c) providers and users of AI systems that are located in a third country, where the output produced by the system is used in the Union.[19]

But policy makers should only use the term AI to define the scope of application if they can also define it in a way that is appropriate for regulatory purposes. The question is: does such a definition exist? To answer this question, I propose a set of requirements for legal definitions generally, survey existing AI definitions, and then discuss the extent to which they meet the requirements for legal definitions.

---

[18] Chris Reed, 'How Should We Regulate Artificial Intelligence?' (2018) 376 *Philosophical Transactions of the Royal Society A* 1, https://doi.org/10.1098/rsta.2017.0360, 2; Bryan Casey and Mark A Lemley, 'You Might Be a Robot' (2019) 105 *Cornell Law Review* 287, https://perma.cc/Y989-ZDXG, 288; Miriam C Buiten, 'Towards Intelligent Regulation of Artificial Intelligence' (2019) 10 *European Journal of Risk Regulation* 41, https://doi.org/10.1017/err.2019.8, 45; Urs Gasser and Virgilio AF Almeida, 'A Layered Model for AI Governance' (2017) 21 *IEEE Internet Computing* 58, https://doi.org/10.1109/MIC.2017.4180835.

[19] European Commission, 'Proposal for a Regulation Laying Down Harmonised Rules on Artificial Intelligence (Artificial Intelligence Act)' COM (2021) 206 final, https://perma.cc/V2RH-6KGC, Art. 2(1).

### *2.1. Requirements for legal definitions*

In democratic countries, policy makers are bound by higher-ranking sources of law, such as constitutional law and general legal principles. If regulations violate these laws or principles, they can be void or invalid—the particular effects are of course jurisdiction-specific. Here, I give a brief overview of relevant laws and principles in the EU and US and distil them into a list of requirements for legal definitions (Table 1).

*Table 1:* Requirements for legal definitions

| Title | Description | Origin |
|---|---|---|
| Over-inclusiveness | Legal definitions must not be over-inclusive. A definition is over-inclusive if it includes cases which are not in need of regulation according to the regulation's objective.[20] Simply put, this is a case of too much regulation. | Principle of proportionality |
| Under-inclusiveness | Legal definitions must not be under-inclusive. A definition is under-inclusive if cases which should have been included are not included.[21] This is a case of too little regulation. | Effectiveness |
| Precision | Legal definitions must be precise. It must be possible to determine clearly whether or not a particular case falls under the definition. | Principle of legal certainty, vagueness doctrine |
| Understandability | Legal definitions must be understandable. Ideally, the definition should be based on the existing meaning of terms and comply with the natural use of language. At least in principle, people without expert knowledge should be able to apply the definition. | Principle of legal certainty, vagueness doctrine |
| Practicability | Legal definitions should be practicable. It should be possible to determine with little effort whether or not a concrete case falls under the definition. The assessment of every element of the definition should be possible on the basis of the information typically available to legal practitioners. | Good legislative practice (helps to maintain the efficiency of the judicial system) |
| Flexibility | Legal definitions should be flexible. They should be able to accommodate technical progress. They should only contain elements which are unlikely to change in the foreseeable future. | Good legislative practice (helps to prevent the need for regulatory updating) |

[20] Robert Baldwin, Martin Cave and Martin Lodge, *Understanding Regulation: Theory, Strategy, and Practice* (Oxford University Press 2011), 70.

[21] Ibid.

Regulations in the EU must comply with the *principle of proportionality*. Pursuant to Article 5(4) of the Treaty on European Union, 'the content and form of Union action shall not exceed what is necessary to achieve the objectives of the Treaties.' Although proportionality has not been used as a general principle of constitutional law in the US, it has nonetheless been recognized as an element of constitutional doctrine in several areas of contemporary constitutional law.[22]

EU regulations must further comply with the *principle of legal certainty*. According to the Court of Justice of the European Union, policy makers are required to ensure 'that Community rules enable those concerned to know precisely the extent of the obligations which are imposed on them. Individuals must be able to ascertain unequivocally what their rights and obligations are and take steps accordingly.'[23]

The US *vagueness doctrine*, which is rooted in due process considerations, has similar implications. According to the US Supreme Court, 'a statute which either forbids or requires the doing of an act in terms so vague that men of common intelligence must necessarily guess at its meaning and differ as to its application violates the first essential of due process of law.'[24] Put differently, 'legal protection requires that texts intended in the first place for use by lawyers should be easily understandable by every citizen.'[25]

Finally, regulations should be *effective*. Here, effectiveness refers to the degree to which a given regulation achieves or progresses towards its objectives. It is worth noting that the concept of effectiveness is highly controversial within legal research,[26] but for the purposes of this paper, the debate has no relevant implications.

To the best of my knowledge, a list similar to Table 1 does not currently exist. Existing lists of requirements for AI definitions[27] and scientific definitions in general[28] do not take a legal perspective. And although most of the above mentioned requirements have been discussed in legal scholarship,[29] there seems to be no comprehensive discussion of all requirements. As mentioned above, different scholars focus on different requirements, which tend to be treated as something given and are rarely, if ever, linked to their legal origin.

---

[22] Vicki C Jackson, 'Constitutional Law in an Age of Proportionality' (2015) 124 *Yale Law Journal* 2680, https://perma.cc/B7HB-5NW4, 3104.

[23] Case C-345/06, *Gottfried Heinrich* (2009) ECR I-01659, https://perma.cc/6YML-D4BW.

[24] *Connally v. General Construction Co.* (1926) 269 US 385, https://perma.cc/C2WR-9H2Q.

[25] Heikki ES Mattila, *Comparative Legal Linguistics: Language of Law, Latin and Modern Lingua Francas* (Routledge 2013), 46; Jeanne Price, 'Wagging, Not Barking: Statutory Definitions' (2013) 60 *Cleveland State Law Review* 999, https://perma.cc/VAH7-YNBP, 1031.

[26] See Maria De Benedetto, 'Effective Law from a Regulatory and Administrative Law Perspective' (2018) 9 *European Journal of Risk Regulation* 391, https://doi.org/10.1017/err.2018.52.

[27] Pei Wang, 'On Defining Artificial Intelligence' (2019) 10 *Journal of Artificial General Intelligence* 1, https://doi.org/10.2478/jagi-2019-0002, 3-6.

[28] Rudolf Carnap, *Logical Foundations of Probability* (University of Chicago Press 1950), https://perma.cc/QE4G-YAZ5, 7.

It is worth noting that the list of requirements should be taken with a grain of salt for two reasons. First, this discussion of the legal origins considers only EU and US laws and principles. Consideration of other jurisdictions was beyond the scope of this paper. However, since the underlying rationale is often not jurisdiction-specific, I expect the list to be useful in other jurisdictions as well. Second, this list is unlikely to be exhaustive. There will likely be further requirements in certain jurisdictions. Similarly, some of the requirements might not be as relevant in some jurisdictions as they are in others, or they might take a slightly different form. For example, it seems plausible that different applications of proportionality analysis lead to different interpretations of over-inclusiveness.[30] But these variations seem to be a necessary consequence of my attempt to define requirements that are relevant for policy makers worldwide. In any case, the requirements can be used to evaluate existing definitions of AI and can be adapted to the requirements of different jurisdictions.

---

[29] The problem of over- and under-inclusive AI definitions is discussed by Lyria B Moses, 'Recurring Dilemmas: The Law's Race to Keep up with Technological Change' (2007) 2 *Journal of Law, Technology & Policy* 239, https://perma.cc/4EKU-RV6J, 260-264; Matthew U Scherer, 'Regulating Artificial Intelligence Systems: Risks, Challenges, Competencies, and Strategies' (2016) 29 *Harvard Journal of Law & Technology* 353, https://perma.cc/2CK2-59EK, 361-362, 373; Chris Reed, 'How Should We Regulate Artificial Intelligence?' (2018) 376 *Philosophical Transactions of the Royal Society A* 1, https://doi.org/10.1098/rsta.2017.0360, 2; Rex Martinez, 'Artificial Intelligence: Distinguishing between Types & Definitions' (2019) 19 *Nevada Law Journal* 1015, https://perma.cc/F8YN-7RKZ, 1038; Bryan Casey and Mark A Lemley, 'You Might Be a Robot' (2019) 105 *Cornell Law Review* 287, https://perma.cc/Y989-ZDXG, 325, 327-328; Miriam C Buiten, 'Towards Intelligent Regulation of Artificial Intelligence' (2019) 10 *European Journal of Risk Regulation* 41, https://doi.org/10.1017/err.2019.8, 45. Precision and understandability are addressed by Matthew U Scherer, 'Regulating Artificial Intelligence Systems: Risks, Challenges, Competencies, and Strategies' (2016) 29 *Harvard Journal of Law & Technology* 353, https://perma.cc/2CK2-59EK, 373; Rex Martinez, 'Artificial Intelligence: Distinguishing between Types & Definitions' (2019) 19 *Nevada Law Journal* 1015, https://perma.cc/F8YN-7RKZ, 1035; and flexibility by Lyria B Moses, 'Recurring Dilemmas: The Law's Race to Keep up with Technological Change' (2007) 2 *Journal of Law, Technology & Policy* 239, https://perma.cc/4EKU-RV6J; Rex Martinez, 'Artificial Intelligence: Distinguishing between Types & Definitions' (2019) 19 *Nevada Law Journal* 1015, https://perma.cc/F8YN-7RKZ, 1017; Bryan Casey and Mark A Lemley, 'You Might Be a Robot' (2019) 105 *Cornell Law Review* 287, https://perma.cc/Y989-ZDXG, 357.

[30] See Vicki C Jackson, 'Constitutional Law in an Age of Proportionality' (2015) 124 *Yale Law Journal* 2680, https://perma.cc/B7HB-5NW4.

### *2.2. Existing definitions of AI*

There is no generally accepted definition of the term AI. Since its first usage by McCarthy et al.,[31] a vast spectrum of definitions has emerged. Below, I provide an overview of existing AI definitions. A more comprehensive collection of definitions can be found in relevant literature.[32] Categorizations of different AI definitions have been proposed by Russell and Norvig,[33] Wang,[34] and Bhatnagar et al.[35] The OECD has also published a Framework for the Classification of AI Systems, which is explicitly targeted at policy makers.[36]

The following list contains popular AI definitions which have been proposed by computer scientists and philosophers:

> The science of making machines do things that would require intelligence if done by men.[37]
>
> The art of creating machines that perform functions that require intelligence when performed by people.[38]
>
> The science and engineering of making intelligent machines, especially intelligent computer programs … Intelligence is the computational part of the ability to achieve goals in the world.[39]
>
> That activity devoted to making machines intelligent, and intelligence is that quality that enables an entity to function appropriately and with foresight in its environment.[40]
>
> The study of agents that receive percepts from the environment and perform actions.[41]

Some legal scholars have also proposed definitions of AI:[42]

---

[31] John McCarthy and others, ‘A Proposal for the Dartmouth Summer Research Project on Artificial Intelligence’ (1955) https://perma.cc/S9DU-GWFF.

[32] Shane Legg and Marcus Hutter, ‘A Collection of Definitions of Intelligence’ (2007) https://arxiv.org/abs/0706.3639; Sofia Samoili and others, ‘AI Watch: Defining Artificial Intelligence’ (*European Commission*, 2020) https://doi.org/10.2760/382730.

[33] Stuart Russell and Peter Norvig, *Artificial Intelligence: A Modern Approach* (Pearson 2020).

[34] Pei Wang, ‘On Defining Artificial Intelligence’ (2019) 10 *Journal of Artificial General Intelligence* 1, https://doi.org/10.2478/jagi-2019-0002.

[35] Sankalp Bhatnagar and others, ‘Mapping Intelligence: Requirements and Possibilities’ in Vincent C Müller (ed), *Philosophy and Theory of Artificial Intelligence* (Springer 2018), https://doi.org/10.1007/978-3-319-96448-5_13.

[36] OECD, ‘Framework for the Classification of AI Systems’ (2022) https://doi.org/10.1787/cb6d9eca-en.

[37] Marvin Minsky, *Semantic Information Processing* (MIT Press 1969), v.

[38] Ray Kurzweil, *The Age of Intelligent Machines* (MIT Press 1990), 14.

[39] John McCarthy, ‘What is Artificial Intelligence?’ (12 November 2007) https://perma.cc/QL9Y-AY8A, 2.

[40] Nils J Nilsson, *The Quest for Artificial Intelligence: A History of Ideas and Achievements* (Cambridge University Press 2009), https://perma.cc/CQV7-N233, xiii.

[41] Stuart Russell and Peter Norvig, *Artificial Intelligence: A Modern Approach* (Pearson 2020), vii.

[42] It is worth noting that none of the definitions is intended to be used to define the scope of AI regulations. Scherer only wants to ‘discuss the definitional problems that regulators will have to confront’ (p. 359), while Turner’s definition is meant as a ‘core definition which

Machines that are capable of performing tasks that, if performed by a human, would be said to require intelligence.[43]

The ability of a non-natural entity to make choices by an evaluative process.[44]

A system, program, software, or algorithm that acts autonomously to think rationally, think humanely, act rationally, act humanely, make decisions, or provide outputs.[45]

AI definitions in policy proposals are particularly relevant for this paper:

Software that is developed with one or more of the techniques and approaches listed in Annex I and can, for a given set of human-defined objectives, generate outputs such as content, predictions, recommendations, or decisions influencing the environments they interact with.[46]

(1) Any artificial system that performs tasks under varying and unpredictable circumstances without significant human oversight, or that can learn from experience and improve performance when exposed to data sets. (2) An artificial system developed in computer software, physical hardware, or another context that solves tasks requiring human-like perception, cognition, planning, learning, communication, or physical action. (3) An artificial system designed to think or act like a human, including cognitive architectures and neural networks. (4) A set of techniques, including machine learning, that is designed to approximate a cognitive task. (5) An artificial system designed to act rationally, including an intelligent software agent or embodied robot that achieves goals using perception, planning, reasoning, learning, communicating, decision-making, and acting.[47]

A machine-based system that can, for a given set of human-defined objectives, make predictions, recommendations, or decisions influencing real or virtual environments.[48]

The use of digital technology to create systems capable of performing tasks commonly thought to require intelligence.[49]

It is worth highlighting a few characteristics of these definitions before continuing with the legal analysis. For example, some of the proposed definitions refer to disciplines ('the science of', 'the art of', 'the study of') and others to

---

captures the essence of a term, without delimiting its precise boundaries' (p. 21), and Martinez acknowledges that his definition 'is going to be under- or over-inclusive' (p. 1038).

[43] Matthew U Scherer, 'Regulating Artificial Intelligence Systems: Risks, Challenges, Competencies, and Strategies' (2016) 29 *Harvard Journal of Law & Technology* 353, https://perma.cc/2CK2-59EK, 362.

[44] Jacob Turner, *Robot Rules: Regulating Artificial Intelligence* (Palgrave Macmillan 2019) https://doi.org/10.1007/978-3-319-96235-1, 16.

[45] Rex Martinez, 'Artificial Intelligence: Distinguishing between Types & Definitions' (2019) 19 *Nevada Law Journal* 1015, https://perma.cc/F8YN-7RKZ, 1038.

[46] European Commission, 'Proposal for a Regulation Laying Down Harmonised Rules on Artificial Intelligence (Artificial Intelligence Act)' COM (2021) 206 final, https://perma.cc/V2RH-6KGC, Art. 3(1).

[47] Section 238(g) of the FY2019 National Defense Authorization Act; also used by Russell T Vought, 'Guidance for Regulation of Artificial Intelligence Applications' (*The White House*, 17 November 2020) https://perma.cc/U2V3-LGV6, 1.

[48] OECD, 'Recommendation of the Council on Artificial Intelligence' (2019) OECD/LEGAL/0449, https://perma.cc/M6Z7-BESV, 7

[49] Office for AI, 'A Guide to Using Artificial Intelligence in the Public Sector' (2019) https://perma.cc/8XQU-LRNB, 6.

systems ('software system', 'artificial system', 'machine-based system'). Most serve academic purposes, while only a few are intended to be used in regulations. One might therefore be tempted to only focus on the definitions by policy makers; however, these definitions are often inspired by academic definitions—for example, the definition in Section 238(g) of the FY2019 National Defense Authorization Act is heavily influenced by Russell and Norvig[50]—thus it seems worthwhile to discuss a wider range of definitions.

### *2.3. Do existing AI definitions meet the requirements for legal definitions?*

As outlined above, legal definitions must meet a number of requirements that can be derived from prior-ranking law, or are at least considered good legislative practice. In Table 2, I discuss the extent to which existing AI definitions meet these requirements using the evaluation options 'Yes', 'No', 'Debatable', and 'Unknown'. Although these options give the false impression that the requirements are binary, they are used for convenience. Since courts ultimately have to make yes-or-no decisions (e.g. whether or not a provision is proportionate), this simplification seems acceptable. It goes without saying that the evaluation is necessarily subjective.

*Table 2:* Do existing AI definitions meet the requirements for legal definitions?

| Requirements | Existing definitions of AI |
|---|---|
| Over-inclusiveness | *No.* Existing AI definitions are highly over-inclusive. For example, many systems that are able to achieve goals in the world are clearly not in need of regulation (e.g. game-playing agents). The same holds true for systems that can, for a given set of human-defined objectives, generate outputs that influence their environment. |
| Under-inclusiveness | *No.* Some AI definitions are also under-inclusive. For example, systems which do not achieve their goals—like an autonomous vehicle that is unable to reliably identify pedestrians—would be excluded, even though they can pose significant risks.[51] Similarly, the Turing test[52] excludes systems that do not communicate in natural language, even though such systems may need regulation (e.g. autonomous vehicles). |
| Precision | *No.* Existing AI definitions are highly vague. Many of them define AI in comparison to human intelligence, even though it is highly disputed how |

[50] Stuart Russell and Peter Norvig, *Artificial Intelligence: A Modern Approach* (Pearson 2020), 1-5.

[51] Matthew U Scherer, 'Regulating Artificial Intelligence Systems: Risks, Challenges, Competencies, and Strategies' (2016) 29 *Harvard Journal of Law & Technology* 353, https://perma.cc/2CK2-59EK, 362.

[52] Alan M Turing, 'Computing Machinery and Intelligence' (1950) 59 *Mind* 433, https://doi.org/10.1093/mind/LIX.236.433.

| | |
|---|---|
| | human intelligence should be defined.[53] Other definitions simply replace one difficult-to-define term ('intelligence') with another ('goal').[54] Russell and Norvig's rational agent definition[55] is equally vague, especially with regards to its notion of limited rationality. In complex environments, agents are often unable to take the optimal action. It is therefore sufficient if they take the action that is optimal *in expectation*. However, in many cases, it is impossible to determine ex-ante whether or not a concrete action is expected to be optimal because ground truth is unattainable. Even if it were, no system can always select the optimal action. How often does a system need to take the optimal action in order to be considered rational? |
| Understandability | *Debatable.* It is debatable whether existing definitions are understandable. The term seems intuitive at first glance—it is simply a compound of two commonly used terms: 'artificial' and 'intelligence'. However, as mentioned above, it is far from obvious what intelligence actually means. The intuitive meaning may also be misleading. Due to pop-cultural illustrations of AI, people might anthropomorphize AI.[56] |
| Practicability | *Debatable.* The practicability of many definitions is also debatable. It may be possible to determine whether or not a system is able to achieve its goals on the basis of typically available information. The Turing test,[57] however, would be highly impracticable. Courts would not be able to conduct the test every time they have to decide whether or not a system is considered AI by the law. |
| Flexibility | *Yes.* The definitions seem sufficiently flexible. The fact that some of them are decades old suggests that they can accommodate technical progress. They also seem relatively general and technology-neutral. One could argue that the so-called 'AI effect' speaks against their flexibility. As McCarthy puts it: 'as soon as it works, no one calls it AI any more'.[58] However, this effect only applies to what is generally considered to be AI. It does not necessarily provide a counterargument against the flexibility of specific definitions. |

[53] Shane Legg and Marcus Hutter, 'A Collection of Definitions of Intelligence' (2007) https://arxiv.org/abs/0706.3639.

[54] Matthew U Scherer, 'Regulating Artificial Intelligence Systems: Risks, Challenges, Competencies, and Strategies' (2016) 29 *Harvard Journal of Law & Technology* 353, https://perma.cc/2CK2-59EK, 361.

[55] Stuart Russell and Peter Norvig, *Artificial Intelligence: A Modern Approach* (Pearson 2020).

[56] Arleen Salles, Kathinka Evers and Michele Farisco, 'Anthropomorphism in AI' (2020) 11 *AJOB Neuroscience* 88, https://doi.org/10.1080/21507740.2020.1740350; Bryan Casey and Mark A Lemley, 'You Might Be a Robot' (2019) 105 *Cornell Law Review* 287, https://perma.cc/Y989-ZDXG, 353-355.

[57] Alan M Turing, 'Computing Machinery and Intelligence' (1950) 59 *Mind* 433, https://doi.org/10.1093/mind/LIX.236.433.

[58] Bertrand Meyer, 'John McCarthy' (*Communications of the ACM*, 28 October 2011) https://perma.cc/49S8-3GM6.

Taken together, existing definitions of AI do not meet the most important requirements for legal definitions. They are highly over-inclusive and vague, while their understandability and practicability are debatable. I doubt that there even is a definition which meets all of the requirements. I would argue that definitions of the term AI are inherently over-inclusive and vague. Due to its broadness, the term will always include many different systems with very different risk profiles which must be treated differently. 'We just need a better definition' would therefore be the wrong conclusion. Relatedly, it would be wrong to deploy a social definition of AI, according to which 'AI is what people generally consider to be AI'.[59] Such a definition would not only be circular, it would also not meet the requirements for legal definitions, as it is inherently vague.

One might object that vagueness is an inherent property of many legal definitions.[60] Many laws use imprecise language, but courts have been able to deal with it. Why should the term AI be any different? My response to this objection is twofold. First, vagueness is a matter of degree. It would be wrong to assume that, simply because courts have been able to deal with imprecise language in the past, policy makers can ignore the issue completely. It might be necessary to use terms that are somewhat imprecise, but I would argue that the term AI is close to the edge of the vagueness spectrum. Second, even if policy makers used a single definition of AI, the above mentioned problems would simply be deferred to the judiciary. Courts would have to develop a casuistry which would also have to meet the requirements detailed above. This would not change the nature of the problem, only the actor who has to solve it.

One might insist that the judiciary would in fact be better suited to develop a precise definition of AI.[61] I do not argue against this claim, as it seems to be a matter of legal tradition. Scholars from civil law countries (like me) tend to favour statutory definitions, while common law scholars are more used to definitions developed by courts.

Finally, one might point out that the proposed AI Act does use a single definition of AI.[62] Am I really suggesting that the proposal does not meet the requirements for legal definitions? Again, my response would be twofold. First, I would argue that their definition of AI mostly serves symbolic purposes. The substance lies in Annex I, which contains a list of technical approaches, and Annex III, which contains a list of high-risk applications. In other words, the material scope is only superficially defined by the term AI. Upon closer examination, the term is an 'empty shell'.[63] Overall, their approach is similar to the one I will suggest below. Second, the European Commission was well aware of the above mentioned requirements. The fact that they explain at length why their approach is future-proof, proportionate, and increases legal certainty[64] suggests that, in their view, other approaches might not meet these requirements.

---

[59] Peter Cihon and others, 'Corporate Governance of Artificial Intelligence in the Public Interest' (2021) 12 *Information*, https://doi.org/10.3390/info12070275.

[60] Matthew U Scherer, 'Regulating Artificial Intelligence Systems: Risks, Challenges, Competencies, and Strategies' (2016) 29 *Harvard Journal of Law & Technology* 353, https://perma.cc/2CK2-59EK, 373.

[61] Bryan Casey and Mark A Lemley, 'You Might Be a Robot' (2019) 105 *Cornell Law Review* 287, https://perma.cc/Y989-ZDXG, 341-344; Jacob Turner, *Robot Rules: Regulating Artificial Intelligence* (Palgrave Macmillan 2019) https://doi.org/10.1007/978-3-319-96235-1, 21.

[62] European Commission, 'Proposal for a Regulation Laying Down Harmonised Rules on Artificial Intelligence (Artificial Intelligence Act)' COM (2021) 206 final, https://perma.cc/V2RH-6KGC, Art. 3(1).

In summary, the results of my discussion seem defensible against plausible objections. I therefore recommend that the material scope should not rely on the term AI. Having said that, I do believe that there is value in using the term for communication purposes. For example, policy makers can still call it 'AI regulation'. They might even use the term to define the material scope, as long as it does not play a substantive role.[65]

## 3. What should they do instead?

For the substance of the scope definition, policy makers should take a risk-based approach. Risk-based regulation tries to achieve policy objectives by targeting activities that pose the highest risk, while leaving lower-risk activities unencumbered.[66] The scope of such regulations is defined by the risks they want to address. As Turner puts it, policy makers should not ask 'what is AI?', but 'why do we need to define AI at all?', and 'what is the unique factor of AI that needs regulation?'[67] Or in the words of Casey and Lemley: 'We don't need rules that decide whether a car with certain autonomous features is or is not a

---

[63] It is worth noting that the term is not completely 'empty'. For example, the fact that doing Bayesian statistics on paper is not covered by the scope is because the AI definition in Art. 3(1) requires an AI system to be software.

[64] Ibid, 3, 7, 10.

[65] If the term is indeed used for the scope definition, it is important that the corresponding definition of AI is very broad and does not exclude relevant systems. Many of the above-mentioned definitions seem to meet these requirements.

[66] See Robert Baldwin and Julia Black, 'Driving Priorities in Risk-based Regulation: What's the Problem?' (2016) 43 *Journal of Law and Society* 565, https://doi.org/10.1111/jols.12003, 565.

[67] Jacob Turner, *Robot Rules: Regulating Artificial Intelligence* (Palgrave Macmillan 2019), https://doi.org/10.1007/978-3-319-96235-1, 8, 15.

robot. What we actually need are rules that regulate unsafe driving behaviour.'[68]

This approach is in line with existing policy proposals. For example, in their proposal for an AI Act, the European Commission focuses on high-risk applications, with almost no requirements for systems with low or minimal risk.[69] They also report that most of the respondents to their stakeholder consultation were explicitly in favour of a risk-based approach.[70] Similarly, the German Data Ethics Commission proposes a pyramid of five levels of criticality.[71]

There is an extensive body of literature on risks from AI. Risks have been conceptualised as accident risks,[72] misuse risks,[73] and structural risks.[74] One could also distinguish between near-term and long-term risks, but some scholars have argued convincingly that this distinction is not always useful, mainly because many ethics and safety issues span different time horizons.[75]

There has also been some work on AI risk factors, broadly defined as all factors that contribute to risks from AI. Most notably, Hernández-Orallo et al. have conducted a survey of known safety-relevant characteristics of AI.[76] They distinguish between (1) internal characteristics (e.g. interpretability), (2) effect of the external environment on the system (e.g. the ability of the operator to intervene during operation), and (3) effect of the system on the external environment (e.g. whether the system influences a safety-critical setting).

---

[68] Bryan Casey and Mark A Lemley, 'You Might Be a Robot' (2019) 105 *Cornell Law Review* 287, https://perma.cc/Y989-ZDXG, 342-343.

[69] European Commission, 'Proposal for a Regulation Laying Down Harmonised Rules on Artificial Intelligence (Artificial Intelligence Act)' COM (2021) 206 final, https://perma.cc/V2RH-6KGC, 12.

[70] Ibid, 8.

[71] German Data Ethics Commission, 'Opinion of the Data Ethics Commission' (2019) https://perma.cc/23QM-JNLJ, 177.

[72] Dario Amodei and others, 'Concrete Problems in AI Safety' (2016) https://arxiv.org/abs/1606.06565; Zachary Arnold and Helen Toner, 'AI Accidents: An Emerging Threat' (*Center for Security and Emerging Technology*, July 2021) https://doi.org/10.51593/20200072.

[73] Miles Brundage and others, 'Malicious Use of Artificial Intelligence: Forecasting, Prevention, and Mitigation' (2018) https://arxiv.org/abs/1802.07228.

[74] Remco Zwetsloot and Allan Dafoe, 'Thinking About Risks from AI: Accidents, Misuse and Structure' (*Lawfare*, 11 February 2019) https://perma.cc/7S3K-6L4U.

[75] Seth D Baum, 'Reconciliation between Factions Focused on Near-Term and Long-Term Artificial Intelligence' (2018) 33 *AI & Society* 565, https://doi.org/10.1007/s00146-017-0734-3; Stephen Cave and Seán ÓhÉigeartaigh, 'Bridging Near- and Long-Term Concerns About AI' (2019) 1 *Nature Machine Intelligence* 5, https://doi.org/10.1038/s42256-018-0003-2; Carina Prunkl and Jess Whittlestone, 'Beyond Near- and Long-Term: Towards a Clearer Account of Research Priorities in AI Ethics and Society' (2020) *Proceedings of the AAAI/ACM Conference on AI, Ethics, and Society* 138, https://doi.org/10.1145/3375627.3375803.

[76] José Hernández-Orallo and others, 'Surveying Safety-relevant AI Characteristics' (2019) *Proceedings of the AAAI Workshop on Artificial Intelligence Safety 2019 co-located with the Thirty-Third AAAI Conference on Artificial Intelligence 2019* 57, https://perma.cc/HDS9-3LA2.

Although their categorization is convincing, I do not use it below, mainly because it serves a different purpose. Theirs is intended to reveal neglected areas of research and to suggest design choices for reducing certain safety concerns, whereas I am interested in defining sources of AI risk in a way that meets the requirements for legal definitions. Their categorization also excludes risks caused by 'the malicious or careless use of a correctly-functioning system', which would be relevant in a regulatory context. For similar reasons, I also do not use the categorization by Burden and Hernández-Orallo.[77]

Instead, I use my own simple categorization of the main sources of risks from AI. I distinguish between (1) technical approaches ('how it is made'), (2) applications ('what it is used for'), and (3) capabilities ('what it can do').[78] In the following, I explain each of the three categories along with examples and discuss the extent to which they meet the requirements for legal definitions.

### *3.1. Technical approaches*

Some AI risks are directly linked to certain technical approaches. One such approach is *reinforcement learning*, which is used in games,[79] robotics,[80] recommender systems[81] and nuclear fusion reactors.[82] But using this approach poses a number of inherent risks. For example, if the objective function of a reinforcement learning agent contains explicit specifications only regarding the main goal, it might implicitly express indifference towards other aspects of the environment. This can lead to situations where the agent disturbs its environment in negative ways while pursuing its main goal. This problem is typically referred to as 'negative side effects'.[83] Another problem is 'reward hacking', the exploitation of unintended loopholes in the reward function.[84] A third problem is how we can ensure that agents can be safely interrupted at any time.[85] Policy makers who want to address these risks could use the following definition:

> 'Reinforcement learning' means the machine learning task of learning a policy from reward signals that maximises a value function.[86]

---

[77] John Burden and José Hernández-Orallo, 'Exploring AI Safety in Degrees: Generality, Capability and Control' (2020) *Proceedings of the Workshop on Artificial Intelligence Safety co-located with 34th AAAI Conference on Artificial Intelligence* 36, https://perma.cc/98QU-FVBM.

[78] Note that the categorisation is not intended to be mutually exclusive. As I will discuss below, I recommend using elements of multiple categories to narrow down the scope. The list is probably also not exhaustive, although I do believe that it captures the vast majority of relevant sources of risks.

[79] Julian Schrittwieser and others, 'Mastering Atari, Go, Chess and Shogi by Planning with a Learned Model' (2020) 588 *Nature* 604, https://doi.org/10.1038/s41586-020-03051-4.

[80] Julian Ibarz and others, 'How to Train Your Robot with Deep Reinforcement Learning: Lessons We have Learned' (2020) 14 *The International Journal of Robotics Research* 698, https://doi.org/10.1177/0278364920987859.

[81] M Medhi Afsar, Trafford Crump and Behrouz Far, 'Reinforcement Learning Based Recommender Systems: A Survey' (2021) https://arxiv.org/abs/2101.06286.

[82] Jonas Degrave and others, 'Magnetic Control of Tokamak Plasmas through Deep Reinforcement Learning' (2022) 602 *Nature* 414, https://doi.org/10.1038/s41586-021-04301-9.

Policy makers could also use the terms *supervised learning* and *unsupervised learning* to define the material scope of AI regulations. These approaches are used in a wide range of different systems, including systems that support judicial decision-making[87] or select employees.[88] However, both approaches can lead to discrimination by reproducing biases contained in the training data.[89] They can be defined as follows:

> 'Supervised learning' means the machine learning task of learning a function that maps from an input to an output based on labelled input-output pairs.[90]
>
> 'Unsupervised learning' means the machine learning task of learning patterns in an input even though no explicit feedback is supplied.[91]

Although it can be important to specify certain technical approaches, they should usually not be the main element of the scope definition. I expect them to be more relevant at lower levels of abstractions, assuming that vague provisions are specified in guidelines or standards.

---

[83] Dario Amodei and others, 'Concrete Problems in AI Safety' (2016) https://arxiv.org/abs/1606.06565, 4-7; Victoria Krakovna and others, 'Penalizing Side Effects Using Stepwise Relative Reachability' (2018) https://arxiv.org/abs/1806.01186.

[84] Jack Clark and Dario Amodei, 'Faulty Reward Functions in the Wild' (*OpenAI*, 21 December 2016) https://perma.cc/6HAB-4BWZ.

[85] Laurent Orseau and Stuart Armstrong, 'Safely Interruptible Agents' (*Machine Intelligence Research Institute*, 28 October 2016) https://perma.cc/RYV9-34QL.

[86] Richard S Sutton and Andrew G Barto, *Reinforcement Learning: An Introduction* (MIT Press 2018), 6.

[87] Julia Angwin and others, 'Machine Bias' (*ProPublica*, 23 May 2016) https://perma.cc/KA6N-WG37.

[88] Jeffrey Dastin, 'Amazon Scraps Secret AI Recruiting Tool That Showed Bias Against Women' (*Reuters*, 11 October 2018) https://perma.cc/CMT4-L468.

[89] Tolga Bolukbasi and others, 'Man is to Computer Programmer as Woman is to Homemaker? Debiasing Word Embeddings' (2016) *Proceedings of the 30th International Conference on Neural Information Processing Systems* 4356, https://perma.cc/V6PD-TN5Z; Joy Buolamwini and Timnit Gebru, 'Gender Shades: Intersectional Accuracy Disparities in Commercial Gender Classification' (2018) *Proceedings of the 1st Conference on Fairness, Accountability and Transparency* 77, https://perma.cc/8TEZ-M3GQ.

[90] Stuart Russell and Peter Norvig, *Artificial Intelligence: A Modern Approach* (Pearson 2020), 652-653.

[91] Ibid.

### *3.2. Applications*

Other risks are not linked to technical approaches, but certain applications. By 'application', I mean a system's use-case within its socio-technical context, including what it is used for, how it is used, who uses it, and what their intentions are.[92] Although the concept is a bit fuzzy, we can distinguish between different subcategories of applications, such as a system's general task (e.g. making recommendations or generating content),[93] its sector-specific use-case (e.g. autonomous driving or automated trading), or its role in the deployment process (e.g. whether it is a foundation model[94] or a fine-tuned model).[95]

Autonomous driving is a typical example of an application. Policy makers may want to reduce the risks that autonomous driving poses to road safety and security, physical integrity, and property rights. The material scope of such regulations could be defined using six levels of automation, as described in the technical standard SAE J3016.[96] These definitions have already been adopted by policy makers in the US[97] and the EU.[98]

Policy makers may also want to reduce the specific risks of *facial recognition technology*. A number of studies show that facial recognition technology can have gender or race biases.[99] This is particularly worrying if such systems are used for law enforcement purposes. In the US, some municipalities have therefore started to ban state use of facial recognition technology for law enforcement purposes, including San Francisco[100] and Boston.[101] The European Commission has proposed a similar ban in the EU, with a few narrow exceptions.[102] In addition to discrimination risks, facial recognition also raises severe privacy concerns.[103] Policy makers who want to address these risks could use the following definition:

> 'Facial recognition' means the automatic processing of digital images which contain the faces of individuals for identification, authentication/verification or categorisation of those individuals.[104]

---

[92] This is related to the socio-technical characteristics specified by NIST, 'AI Risk Management Framework: Initial Draft' (2022) https://perma.cc/FGM8-5TTG, 10-12.

[93] I owe this idea to Markus Anderljung.

[94] 'Foundation models' are large pre-trained models that can serve as the foundation for a wide array of down-stream applications. Some predict that their use will be increasingly widespread, Rishi Bommasani and others, 'On the Opportunities and Risks of Foundation Models' (2021) https://arxiv.org/abs/2108.07258.

[95] Note that the subcategories are neither mutually exclusive nor exhaustive.

[96] SAE International, 'Taxonomy and Definitions for Terms Related to Driving Automation Systems for On-Road Motor Vehicles' (2021) J3016_202104, https://perma.cc/7LD6-48VG.

[97] US Department of Transportation, 'Preparing for the Future of Transportation' (2018) https://perma.cc/FPJ3-VELU.

[98] European Commission, 'On the Road to Automated Mobility: An EU Strategy for Mobility of the Future' COM (2018) 283 final, https://perma.cc/5ZXR-YUXD.

[99] Joy Buolamwini and Timnit Gebru, 'Gender Shades: Intersectional Accuracy Disparities in Commercial Gender Classification' (2018) *Proceedings of the 1st Conference on Fairness, Accountability and Transparency* 77, https://perma.cc/8TEZ-M3GQ.

[100] Kate Conger, Richard Fausset and Serge F Kovaleski, 'San Francisco Bans Facial Recognition Technology' (*The New York Times*, 14 May 2019) https://perma.cc/B8KQ-P8WX.

[101] Khari Johnson, 'Boston Bans Facial Recognition Due to Concern About Racial Bias' (*VentureBeat*, 24 June 2020) https://perma.cc/G635-AGCY.

Overall, I expect applications to be the most important element of the scope definition, especially tasks and sectoral use-cases.

### *3.3. Capabilities*

A third category of sources of AI risk is a system's capabilities. For example, policy makers may want to limit the material scope to systems which can *physically interact with their environment* via robotic hands or other actuators. Only embodied systems can directly cause physical harm or damage property.[105] This ability could be defined as follows:

> 'Physical interaction' means the ability to use sensors to perceive the physical environment and effectors to manipulate this environment.[106]

Another capability-related source of AI risk is the *ability to make automated decisions*. This element can be used to exclude systems which only make suggestions while humans make the final decision. One could call systems with this ability 'self-executive'. Policy makers could use this element to address certain risks resulting from a loss of control[107] and other assurance risks—those risks which stem from an operator's inability to understand and control AI systems during operation.[108] This element is already being used in Articles 13(2)(f), 14(2)(g) and 15(1)(h) of the GDPR. It can be defined as follows:

> 'Automated decision-making' means the ability to make decisions by technological means without human involvement.[109]

---

[102] European Commission, 'Proposal for a Regulation Laying Down Harmonised Rules on Artificial Intelligence (Artificial Intelligence Act)' COM (2021) 206 final, https://perma.cc/V2RH-6KGC.

[103] Zekeriya Erkin and others, 'Privacy-Preserving Face Recognition' (2009) *Proceedings of the 9th International Symposium on Privacy Enhancing Technologies* 235, https://doi.org/10.1007/978-3-642-03168-7_14.

[104] Article 29 Data Protection Working Party, 'Opinion 02/2012 on Facial Recognition in Online and Mobile Services' (2012) WP192, https://perma.cc/Y72E-WAX3, 2.

[105] José Hernández-Orallo and others, 'Surveying Safety-relevant AI Characteristics' (2019) *Proceedings of the AAAI Workshop on Artificial Intelligence Safety 2019 co-located with the Thirty-Third AAAI Conference on Artificial Intelligence 2019* 57, https://perma.cc/HDS9-3LA2, 58.

[106] Stuart Russell and Peter Norvig, *Artificial Intelligence: A Modern Approach* (Pearson 2009).

[107] Matthew U Scherer, 'Regulating Artificial Intelligence Systems: Risks, Challenges, Competencies, and Strategies' (2016) 29 *Harvard Journal of Law & Technology* 353, https://perma.cc/2CK2-59EK, 366-369.

A third example of a capability is the *ability to make decisions which have a legal or similarly significant effect*. Consider two virtual assistants: one reminds you on your friends' birthdays, the other is able to buy products. Clearly, the two systems have very different risk profiles (the latter may require some degree of consumer protection, for example). This element is already being used in Article 22 of the GDPR. The European Data Protection Board has endorsed the definition by the Article 29 Data Protection Working Party:[110]

> 'Legal effect' means any impact on a person's legal status or their legal rights.
>
> 'Similarly significant effect' means any equivalent impact on a person's circumstances, behaviour, or choices. This may include their financial circumstances, access to health services, employment opportunities or access to education.

The main role of this class of elements is to narrow down the scope. It should not be the central element.

### *3.4. Do definitions of certain technical approaches, applications, and capabilities meet the requirements for legal definitions?*

Let us now examine to what extent definitions of certain technical approaches, applications, and capabilities meet the requirements for legal definitions. Table 3 breaks down the discussion by category and requirement.

---

[108] Pedro A Ortega and Vishal Maini, 'Building Safe Artificial Intelligence: Specification, Robustness, and Assurance' (*Medium*, 27 September 2018) https://perma.cc/L7PK-LC46.

[109] Article 29 Data Protection Working Party, 'Guidelines on Automated Individual Decision-Making and Profiling for the Purposes of Regulation 2016/679' (2018) WP251rev.01, https://perma.cc/864E-R5MJ, 8.

[110] Ibid, 21-22.

*Table 3:* Do definitions of certain technical approaches, applications, and capabilities meet the requirements for legal definitions?

| Requirements | Technical approaches | Applications | Capabilities |
|---|---|---|---|
| Over-inclusiveness | *No.* There will always be systems that use one of the above mentioned technical approaches, but should not be subject to regulation (e.g. game-playing agents based on reinforcement learning). | *Yes.* In many cases, the main regulatory goal will be to reduce certain application-specific risks (e.g. discriminatory recommender systems used to support judicial decision-making). | *No.* Not all systems with certain capabilities pose risks which are in need of regulation. For example, industrial robots and vending machines both have the ability to physically manipulate their environment, but their risk profile is very different. |
| Under-inclusiveness | *No.* Relevant risks can not be attributed to a single technical approach. For example, supervised learning is not inherently risky. And if a definition lists many technical approaches, it would likely be over-inclusive. | *No.* Not all systems that are applied in a specific context pose the same risks. Many of the risks also depend on the technical approach. | *No.* Relevant risks can not be attributed to a certain capability alone. By its very nature, capabilities need to be combined with other elements (‘capability of something’). |
| Precision | *Yes.* It is easy to determine whether or not a system is based on a certain technical approach. | *Yes.* Applications can be defined precisely. This is by no means a novel challenge for the law. | *Yes.* In many cases, capabilities can be defined in a binary way (e.g. a system either can physically manipulate its environment or not). |
| Understandability | *Yes.* For developers it will be easy to understand definitions of certain technical approaches. One can expect the same from non-technical people who are responsible for the development, deployment, or use of systems. | *Yes.* There are no apparent reasons for why definitions of applications are not understandable. | *Yes.* Most capabilities are intuitive (e.g. the ability to physically manipulate its environment). |
| Practicability | *Yes.* The required information about the technical approach is easy to obtain. | *Yes.* The required information about the application is easy to obtain. | *Yes.* Some capabilities already have established legal definitions (e.g. the ability to make decisions which have a legal or similarly significant effect). |
| Flexibility | *Unknown.* It is highly uncertain whether today's technical approaches will be used in the future. Definitions will be more flexible if the technical approach is defined broadly, but they will also be less precise. | *Debatable.* While some applications are unlikely to change in the future, almost certainly new applications will emerge. | *Yes.* Definitions of capabilities seem to be able to accommodate technical progress. |

In summary, definitions of certain technical approaches, applications, and capabilities meet more of the requirements for legal definition than definitions of the term AI (see Table 2). This suggests that policy makers should favour a risk-based approach over the 'classical' approach.

One might be tempted to simply pick one of three categories for the scope definition, but I would argue that a *multi-element approach* is often preferable.[111] The following example illustrates the idea:

> This regulation applies to facial recognition systems for law enforcement purposes based on supervised learning.

In the example, the material scope is defined by a certain application (facial recognition for law enforcement purposes) and a certain technical approach (supervised learning). This approach allows policy makers to target risks in a more fine-grained way and thereby reduce over-inclusiveness and increase precision.

Relatedly, there will always be cases which fall under the scope definition, but should not be included. To further reduce over-inclusiveness, policy makers can use exemptions.[112] For example, Article 2(3) of the AI Act contains the following exemption:

[111] Bryan Casey and Mark A Lemley, 'You Might Be a Robot' (2019) 105 *Cornell Law Review* 287, https://perma.cc/Y989-ZDXG, 356.

[112] Note that there is a difference between 'not covered by the scope' and 'exempt from scope'. In the first case, the regulation is not applicable. In the second case, the regulation is applicable, but it explicitly states that it should not apply to a specific case.

This Regulation shall not apply to AI systems developed or used exclusively for military purposes.[113]

Exemptions can be located at the beginning of the regulation (e.g. within the definition of the material scope), in the body of the regulation (e.g. within particular chapters or norms), or both. In the context of risk-based AI regulation, exemptions will typically cover cases where negative effects are low-probability, low-impact (scale), where systems only affect a small number of people (scope), and/or where people can decide not to be subject to the effects of the system (optionality).[114] Exemptions may also be used to exclude areas which are already governed by other areas of law (e.g. military use of AI). Overall, I expect most AI regulations to benefit from exemptions in one way or another.

The proposed AI Act takes a similar approach to the one I recommend.[115] As mentioned above, the material scope is not really defined by the term AI. Instead, the scope definition combines a number of technical approaches (Annex I) with certain high-risk applications (Annex III).[116] Although this seems like a reasonable approach, I would point out three potential areas for improvement. First, to the extent that my observation is correct, the European Commission should consider making it explicit that their scope definition does not rely on the term AI (e.g. in the recitals). This could help to prevent misconceptions among laypeople (e.g. the false interpretation that the regulation would apply to any use of Bayesian statistics[117]). Second, they should consider distinguishing between different technical approaches. In the current version, it is sufficient if a system is based on *any* of the technical approaches listed in Annex I. However, a recruiting system based on a simple statistical approach would not pose the same risks as a system based on supervised learning. Third, they should consider defining capabilities, as doing so could further reduce over-inclusiveness and increase precision.

---

[113] European Commission, 'Proposal for a Regulation Laying Down Harmonised Rules on Artificial Intelligence (Artificial Intelligence Act)' COM (2021) 206 final, https://perma.cc/V2RH-6KGC, Art. 2(3).

[114] See OECD, 'Framework for the Classification of AI Systems' (2022) https://doi.org/10.1787/cb6d9eca-en, 67.

[115] European Commission, 'Proposal for a Regulation Laying Down Harmonised Rules on Artificial Intelligence (Artificial Intelligence Act)' COM (2021) 206 final, https://perma.cc/V2RH-6KGC.

[116] The third element—the ability to generate outputs that influence environments—seems to not play any meaningful role.

[117] As implied by Bob Carpenter, 'EU Proposing to Regulate the Use of Bayesian Estimation' (*Statistical Modeling, Causal Inference, and Social Science*, 22 April 2021) https://perma.cc/6FTJ-9FZB.

## 4. Can this approach also be applied to AGI regulations?

Future AI systems that achieve or exceed human performance in a wide range of cognitive tasks have been referred to as 'artificial general intelligence (AGI)'.[118] Even though the prospect of AGI is speculative, and some people remain sceptical,[119] a number of surveys show that many AI researchers do take it seriously.[120]

While the development of AGI could be overwhelmingly beneficial for humanity, it could also pose significant risks. Potential risks from AGI have been studied, among others, by Nick Bostrom,[121] Allan Dafoe,[122] Stuart Russell,[123] and Toby Ord.[124] There are also a number of public figures, such as Stephen Hawking,[125] Elon Musk,[126] and Bill Gates,[127] who have warned against the dangers of AGI. Against this background, it is not surprising that policy makers have started taking AGI more seriously. For example, the UK National AI Strategy contains the following passage:

> The [UK] government takes the long-term risk of non-aligned Artificial General Intelligence, and the unforeseeable changes that it would mean for the UK and the world, seriously.[128]

---

[118] Ben Goertzel and Cassio Pennachin, *Artificial General Intelligence* (Springer 2007), https://doi.org/10.1007/978-3-540-68677-4.

[119] Oren Etzioni, 'No, the Experts Don't Think Superintelligent AI is a Threat to Humanity' (*MIT Technology Review*, 20 September 2016) https://perma.cc/ZME3-UGXW; Melanie Mitchell, 'Why AI is Harder than We Think' (2021) https://arxiv.org/abs/2104.12871.

[120] Seth D Baum, Ben Goertzel and Ted G Goertzel, 'How Long Until Human-Level AI? Results from an Expert Assessment' (2011) 78 *Technological Forecasting and Social Change* 185, https://doi.org/10.1016/j.techfore.2010.09.006; Vincent C Müller and Nick Bostrom, 'Future Progress in Artificial Intelligence: A Survey of Expert Opinion' in Vincent C Müller (ed), *Fundamental Issues of Artificial Intelligence* (Springer 2016), https://doi.org/10.1007/978-3-319-26485-1_33; Katja Grace and others, 'When Will AI Exceed Human Performance? Evidence from AI Experts' (2018) 62 *Journal of Artificial Intelligence Research* 729, https://doi.org/10.1613/jair.1.11222.

[121] Nick Bostrom, *Superintelligence: Paths, Dangers, Strategies* (Oxford University Press 2014).

[122] Allan Dafoe, 'AI Governance: A Research Agenda' (*Centre for the Governance of AI*, 27 August 2018) https://perma.cc/SA6T-F6XW.

[123] Stuart Russell, *Human Compatible: Artificial Intelligence and the Problem of Control* (Penguin Random House 2019).

[124] Toby Ord, *The Precipice: Existential Risk and the Future of Humanity* (Hachette Books 2020).

[125] Rory Cellan-Jones, 'Stephen Hawking Warns Artificial Intelligence Could End Mankind' (*BBC*, 2 December 2014) https://perma.cc/GA8A-BT5M.

[126] Samuel Gibbs, 'Elon Musk: Artificial Intelligence is our Biggest Existential Threat' (*The Guardian*, 27 October 2014) https://perma.cc/8WLM-LN4G.

[127] Kevin Rawlinson, 'Microsoft's Bill Gates Insists AI is a Threat' (*BBC*, 29 January 2015) https://perma.cc/N6Y8-CG2S.

[128] UK Government, 'National AI Strategy' (2021) https://perma.cc/RYN4-EEBR, 60.

If and when it becomes evident that AGI is in fact possible, policy makers may want to reduce the associated risks via regulation. This would again raise the question of how they should define the material scope of such AGI regulations. Would a risk-based approach be applicable to define all sorts of AI, including AGI?

It seems very likely that the *technical approach* that is used to build AGI will significantly influence its risks and potential risk mitigation strategies. For example, if AGI is developed using reinforcement learning,[129] we might use an approach called 'reward modelling' to align it to human values.[130] One might therefore be tempted to rely on technical approaches when defining the material scope of AGI regulations. However, there is an ongoing debate about whether today's technical approaches are sufficient to build AGI. While some AI researchers think this is reasonable,[131] others remain sceptical.[132] Given the high degree of uncertainty, policy makers should probably not rely exclusively on specific technical approaches.

Since AGI is characterised by the generality of its intelligence, it seems less fruitful to define specific *applications*. However, one could nonetheless distinguish between different types of AGI, such as question-answering, command-executing, or non-goal-directed systems.[133] Since these types could influence the feasibility and desirability of different safety precautions,[134] policy makers may want to use them to define the material scope of AGI regulations.

As mentioned above, the decisive *capability* of AGI is the generality of its intelligence. If a system exceeds human intelligence across the board, humanity would become the second most intelligent species on Earth[135] and might permanently lose its influence over the future.[136] However, I doubt that there is a definition of this capability that meets the requirements for legal definitions, mainly because I expect it to be highly vague. Instead, policy makers may want to define capabilities that could lead to the development of AGI.[137] These capabilities seem easier to define, but would still capture relevant AGI risks. One such capability could be the ability to recursively self-improve:[138]

---

[129] David Silver and others, 'Reward is Enough' (2021) 299 *Artificial Intelligence*, https://doi.org/10.1016/j.artint.2021.103535.

[130] Jan Leike and others, 'Scalable Agent Alignment via Reward Modeling: A Research Direction' (2018) https://arxiv.org/abs/1811.07871.

[131] Paul Christiano, 'Prosaic AI Alignment' (*AI Alignment*, 19 November 2016) https://perma.cc/43ED-M735.

[132] Daniel Kokotajlo, 'A Dilemma for Prosaic AI Alignment' (*AI Alignment Forum*, 17 December 2019) https://perma.cc/SG22-SCVW.

[133] Nick Bostrom, *Superintelligence: Paths, Dangers, Strategies* (Oxford University Press 2014), 177-193.

[134] Ibid, 191-192.

[135] Richard Ngo, 'AGI Safety from First Principles' (*AI Alignment Forum*, 28 September 2020) https://perma.cc/8JEE-ZDH8.

[136] Nick Bostrom, *Superintelligence: Paths, Dangers, Strategies* (Oxford University Press 2014); Toby Ord, *The Precipice: Existential Risk and the Future of Humanity* (Hachette Books 2020).

> 'Recursive self-improvement' means an agent's ability to iteratively improve its own performance.

In summary, it seems plausible that policy makers could follow a risk-based approach to define the material scope of AGI regulations, though the focus might shift from technical approaches to capabilities.

A final note on foreseeability: It seems highly unlikely that we are able to foresee all future risks from AI. It is therefore necessary to make the scope definition scalable and adaptable. One possible approach could be to make the substance of the scope definition more future-proof. For example, one could include a catch-all definition of AI risk ('any other instance of significant risk caused or exacerbated by…'), which regulators and courts could then use to fill future regulatory gaps. Another approach would be to make it easier to update the scope definition as we identify new risks (e.g. via sunset clauses or built-in revision schedules). One could also combine more general definitions at the legislative level with more specific definitions at the sub-legislative level—which is very similar to what the proposed AI Act does.[139] Regardless of the particular approach policy makers choose, they need to closely monitor the AI landscape[140] and pay close attention to early warning signs.[141]

## 5. Conclusion

In this paper, I have shown that existing definitions of AI do not meet the most important requirements for legal definitions. Therefore, policy makers should not rely substantially on the term AI to define the material scope of AI. I have also shown that definitions of the main sources of relevant risks—certain technical approaches, applications, and capabilities—meet more of the requirements for legal definitions than definitions of the term AI. Finally, I have argued that this approach can, in principle, also be used to define the material scope of AGI regulations.

---

[137] More specifically, the scope definition could focus on the *intention* to develop certain capabilities (e.g. 'any serious and promising attempt').

[138] Nick Bostrom, *Superintelligence: Paths, Dangers, Strategies* (Oxford University Press 2014), 409; Irving J Good, 'Speculations Concerning the First Ultraintelligent Machine' (1966) 6 *Advances in Computers* 31, https://doi.org/10.1016/S0065-2458(08)60418-0, 33.

[139] Pursuant to Art. 4, the European Commission can adopt delegated acts to amend the list of techniques and approaches listed in Annex I.

[140] Jess Whittlestone and Jack Clark, 'Why and How Governments Should Monitor AI Development' (2021) https://arxiv.org/abs/2108.12427.

[141] Carla Z Cremer and Jess Whittlestone, 'Artificial Canaries: Early Warning Signs for Anticipatory and Democratic Governance of AI' (2020) 6 *International Journal of Interactive Multimedia and Artificial Intelligence* 100, http://doi.org/10.9781/ijimai.2021.02.011.

The paper has made four main contributions. First, it has provided a comprehensive legal argument for why policy makers should not rely on the term AI for regulatory purposes and why a risk-based approach would be preferable. Second, it has proposed a list of specific requirements for legal definitions which can also be used to evaluate other definitions. Third, the paper has suggested a new categorization of the main sources of AI risks that policy makers may want to address. And fourth, it can be seen as a first step towards the study of AGI regulation, which I expect will turn into its own field of interest for policy makers and researchers in the future.

The findings of this paper are relevant for policy makers worldwide. They support the European Commission's risk-based approach.[142] The suggested definitions of certain technical approaches, applications, and capabilities can also be used to amend or substantiate the list of techniques and approaches in Annex I and high-risk applications in Annex III. But I expect the findings to be even more relevant for policy makers who have not yet drafted concrete proposals. Defining the material scope of AI regulations requires careful consideration. I hope this paper comes at the right time to help policy makers rise to this challenge.

## Acknowledgements

I am grateful for valuable comments and feedback from Seth Baum, Conor Griffin, Renan Araújo, Leonie Koessler, Nick Hollman, Suzanne Van Arsdale, Markus Anderljung, Matthijs Maas, and Sébastien Krier. I also thank the participants of a seminar hosted by the Legal Priorities Project in February 2021. All remaining errors are my own.

[142] European Commission, 'Proposal for a Regulation Laying Down Harmonised Rules on Artificial Intelligence (Artificial Intelligence Act)' COM (2021) 206 final, https://perma.cc/V2RH-6KGC.